%% file: conference_101719.tex
\documentclass[conference]{IEEEtran}
\IEEEoverridecommandlockouts
\usepackage{cite}
\usepackage{amsmath,amssymb,amsfonts}
\usepackage{algorithmic}
\usepackage{graphicx}
\usepackage{textcomp}
\usepackage{xcolor}
\usepackage{multirow} 
\usepackage{subcaption}

\newcommand{\ignore}[1]{}

\newcommand{\linebreakand}{%
  \begin{@IEEEauthorhalign}
  \hfill\mbox{}\par
  \mbox{}\hfill \end{@IEEEauthorhalign}
}

\def\BibTeX{{\rm B\kern-.05em{\sc i\kern-.025em b}\kern-.08em
    T\kern-.1667em\lower.7ex\hbox{E}\kern-.125emX}}

\begin{document}

\title{Characterizing Event-themed Malicious Web Campaigns: A Case Study on War-themed Websites}

\author{\IEEEauthorblockN{Maraz Mia}
\IEEEauthorblockA{\textit{Department of Computer Science} \\
\textit{Tennessee Tech University}\\
Cookeville, TN, USA\\
mmia43@tntech.edu}
\and
\IEEEauthorblockN{Mir Mehedi A. Pritom}
\IEEEauthorblockA{\textit{Department of Computer Science} \\
\textit{Tennessee Tech University}\\
Cookeville, TN, USA\\
mpritom@tntech.edu}
\linebreakand

\IEEEauthorblockN{Tariqul Islam}
\IEEEauthorblockA{\textit{School of Information Studies} \\
\textit{Syracuse University}\\
Syracuse, NY, USA \\
mtislam@syr.edu}
\and
\IEEEauthorblockN{Shouhuai Xu}
\IEEEauthorblockA{\textit{Department of Computer Science} \\
\textit{University of Colorado Colorado Springs}\\
Colorado Springs, CO, USA \\
sxu@uccs.edu}
}


\maketitle

\begin{abstract}
Cybercrimes such as online scams and fraud have become prevalent. Cybercriminals often abuse various global or regional events as themes of their fraudulent activities to breach user trust and attain a higher attack success rate. These attacks attempt to manipulate and deceive innocent people into interacting with meticulously crafted websites with malicious payloads, phishing, or fraudulent transactions. 
To deepen our understanding of the problem, this paper investigates how to characterize event-themed malicious website-based campaigns, with a case study on war-themed websites. We find that attackers tailor their attacks by exploiting the unique aspects of events, as evidenced by activities such as fundraising, providing aid, collecting essential supplies, or seeking updated news. We use explainable unsupervised clustering methods to draw further insights, which could guide the design of effective early defenses against various event-themed malicious web campaigns. 
\end{abstract}


\vspace{-0.70em}
\section{Introduction}
\label{sec:intro}
\input{Introduction.tex}

\vspace{-0.50em}
\section{Related Work}
\label{sec:related-works}
\input{related-work}

\vspace{-0.5em}
\section{Methodology}
\label{sec:methodology}
\input{methodology.tex}

\section{Case Study}
\label{sec:casestudy}
\input{studyresults.tex}

\section{Limitations}
\label{sec:discuss}
\input{discussion}


\section{Conclusion}
\label{sec:conclusion}
\input{conclusion}

\bibliographystyle{IEEEtran}
 \bibliography{reference}
 
\end{document}

%% file: Introduction.tex
Event-themed cyber social engineering attacks are an emerging threat, where an {\em event} 
can be any global or regional incidents, such as COVID-19, disasters (e.g., hurricanes, earthquakes), or wars \cite{al2022covid,mir2020characterizing,khatri2023global,rakesh_verma_phishing_with_disaster}.
The exploited events often 
overwhelm the public and/or make people distressed.
Putting other words, events create a large attack surface to attract mischievous groups to carry out online scams and frauds with a massive number of potentially vulnerable users. 
One fairly recent example is the Russia-Ukraine war,
which has been exploited by many online scams 
\cite{ukraine_cyber_scams,ukraine_charity_scams}, highlighting that attackers have exploited the war situation to wage cyber social engineering attacks. These attacks can be stealthy in the sense that it is often nontrivial to distinguish them from legitimate donation campaigns.
This motivates us to investigate the identification of event-themed malicious websites in near real-time.
To the best of our knowledge, there is no systematic study on characterizing event-themed website-based cyber social engineering attacks 
despite numerous studies (cf. \cite{XuPIEEE2024} and the references therein).

\noindent{\bf Contributions}. In this paper, we make two major contributions. First, we initiate the systematic study on event-themed malicious website-based campaigns by proposing a methodology
that leverages unsupervised clustering methods to study event-themed website campaigns. The choice of unsupervised learning can be justified by the fact that the exploited events are often new incidents, meaning that there is no historical data that can be leveraged for supervised learning purposes. 
Moreover, the methodology applies machine learning (ML) explainability to justify why a particular clustering result is obtained. 
Second, we demonstrate the usefulness of the methodology by conducting a case study on Russia-Ukraine war-themed website campaigns. Our findings include: 
(a) clustering is an effective method because malicious websites often belong to a particular cluster, hinting that attackers may use common tactics;
(b) website structure, such as the home page size and the number of pages within a website, plays an important role in characterizing the clusters; 
(c) different clusters exhibit different trends of Top Level Domains (TLDs) usage; and (d) the wavering nature of websites oscillating from parked to normal state may indicate that a website has been compromised and/or waiting for propagating attacks.

\ignore{
\begin{itemize}
     \item Finding out specific indicators of compromise relevant to Russia-Ukraine war-themed malicious websites.
     \item Providing an in-depth analysis of both malicious and benign websites within this themed cyberattacks.
     
     \item Incorporating methods for evaluating/assessing the clustering results with deeper insights about the reasoning (explanation) behind these groups of websites.  
     \item Labeling the identified groups of website clusters as potentially malicious or bad websites. 
 \end{itemize}
}

\noindent{\bf Outline}. 
Section \ref{sec:related-works} discusses related prior studies.
Section \ref{sec:methodology} describes our characterization methodology. Section \ref{sec:casestudy} presents our case study. Section \ref{sec:discuss} discusses the limitations of the present study. 
Section \ref{sec:conclusion} concludes the paper.

%% file: related-work.tex
We divide closely related previous studies into two thrusts, \textit{ event-themed cyber attacks} and, in general, \textit{ malicious websites}, because the present study is rooted in their intersection. 

\noindent{\bf Event-Themed Cyber Attacks}.
Event-themed cyber attacks are a real concern.
At the beginning of the COVID-19 pandemic, Pritom et al. \cite{mir2020characterizing} characterized methods and attack kill chains used by COVID-19-themed cyber attacks across propagation mediums. Research also showed that various cyber social engineering attack techniques have been exploited for gaining trust  \cite{Oest_20_apwgSymp_scamPandemic,XuPIEEE2024}. 
Researchers have characterized COVID-19-themed malicious websites based on lexical and WHOIS features of the website domain names \cite{mir_covid_website20}. 
Behzad et al. \cite{Behzad_2023CovidScams_AsiaCCS} used a clustering approach to characterize COVID-19-themed scam webpages containing malware payloads via their visual similarity of the webpages. However, a manual labeling approach by the authors and the clustering of only scam sites based on visual similarity can be ineffective when there is a mixture of benign and malicious websites. Other researchers have also discussed the challenges with modern technologies during natural disasters and emergency management \cite{krichen2023managing}. However, the aforementioned studies mostly dealt with labeled data, meaning that they worked with the later stages of such attacks because, availability of labeled data would be critical at the initial stage for an ongoing event. This is true despite that event-themed attacks are often in the forms of 
phishing, misinformation, distributed denial of services, ransomware, malware, scam, and website defacement attacks \cite{magafasrussia,mohee2022cyber,vu2022getting}
and the primary targets of cyber attacks during any war were commercial and government networks
\cite{magafasrussia,vu2022getting,gabrian2022russia,gabrian2022russia}, yet the other attack types, which are aimed at the normal user via websites, are still not well explored.

\noindent{\bf Malicious Websites}. 
Malicious websites may or may not be associated with any particular event. 
There are many studies on detecting web-based attacks  (e.g., Phishing, Malware, fraud), typically leveraging machine learning  
(e.g., 
\cite{PhishingURL_data_diversity_Nsys2024_mir,sensors23_malURLs_ml,phish_urls_verma_codaspy15,xu13_crosslayer_malwebsite}),
and on protection methods \cite{ccs21_phishing_https_certificate}.
User interactions within phishing websites have been studied at scale to understand phishing patterns \cite{phishpatterns_IMC22_Perdisci}. Researchers studied {\em adversarial} malicious websites detection (i.e., evasion attacks against machine learning-based detectors) via {\em proactive training} \cite{xu14_cns_evasion_malwebsite} even before the notion of {\em adversarial training} was introduced \cite{goodfellow2015explainingICLR2015} in other contexts. 
Adversarial malicious websites attempt to evade detection by manipulating some features used by detectors
\cite{xu14_cns_evasion_malwebsite,das2019sok_rakesh,song2021advanced}. 
On the other hand, blocklists have long been used by browsers to prevent users from visiting potentially malicious websites. 
Attackers have also nullified this defense by exploiting cloaking techniques \cite{Oest20_Phishtime,CrawlPhish_Oest_sp22_cloaking}. Scammers also introduced new attacks against blocklisting \cite{Oest_Phishfarm_sp19_blacklist}. Other solutions include providing more reliable results on large-scale phishing detection against previously unseen evasive techniques 
\cite{Oest_sp23_BeyondPhish,oest2022neutralizing}. However, none of these studies focus on dealing with malicious website campaigns when the labeled data is absent to train a supervised model. Further, the manual process of labeling or collecting data from different blacklists may provide strategic advantage to the attackers as they can crack the detection mechanism through evasive adaptation. 

%% file: methodology.tex
Event-themed malicious websites are emerging threats, meaning that little information about the attacks is available to the defender. Thus, our methodology is driven by the following research questions (RQs):
\begin{itemize} 

\item{\bf RQ1}.
How can we group (or cluster) event-themed websites based on their similarity in terms of their characteristics? 


\item{\bf RQ2}.
How can we characterize the website groups (e.g., clusters) in terms of website structure (e.g., number of pages per website, home page size), differences among their TLD (Top-Level Domain) usage, and differences in their keyword usage within domain name strings to better understand each group's identity?

\item {\bf RQ3}.
Can the websites that are grouped be tagged as themed malicious campaigns? If so, what are the types (e.g., donation, cause-campaign, services, etc.) and indicators of such campaigns? 




\item {\bf RQ4}.
Are there any particular cluster (or clusters) that are more likely to be identified as {\em malicious}? 
\end{itemize}

To address these RQs, we propose a methodology incorporating the following six modules: (a) problem formalization; 
(b) data collection and pre-processing; (c) feature extraction and selection; (d) unsupervised machine learning or clustering; (e) cluster explanation and (f) cluster characterization. 
    
\begin{figure*}[!htbp]
    \centering
\includegraphics[width=0.75\textwidth]{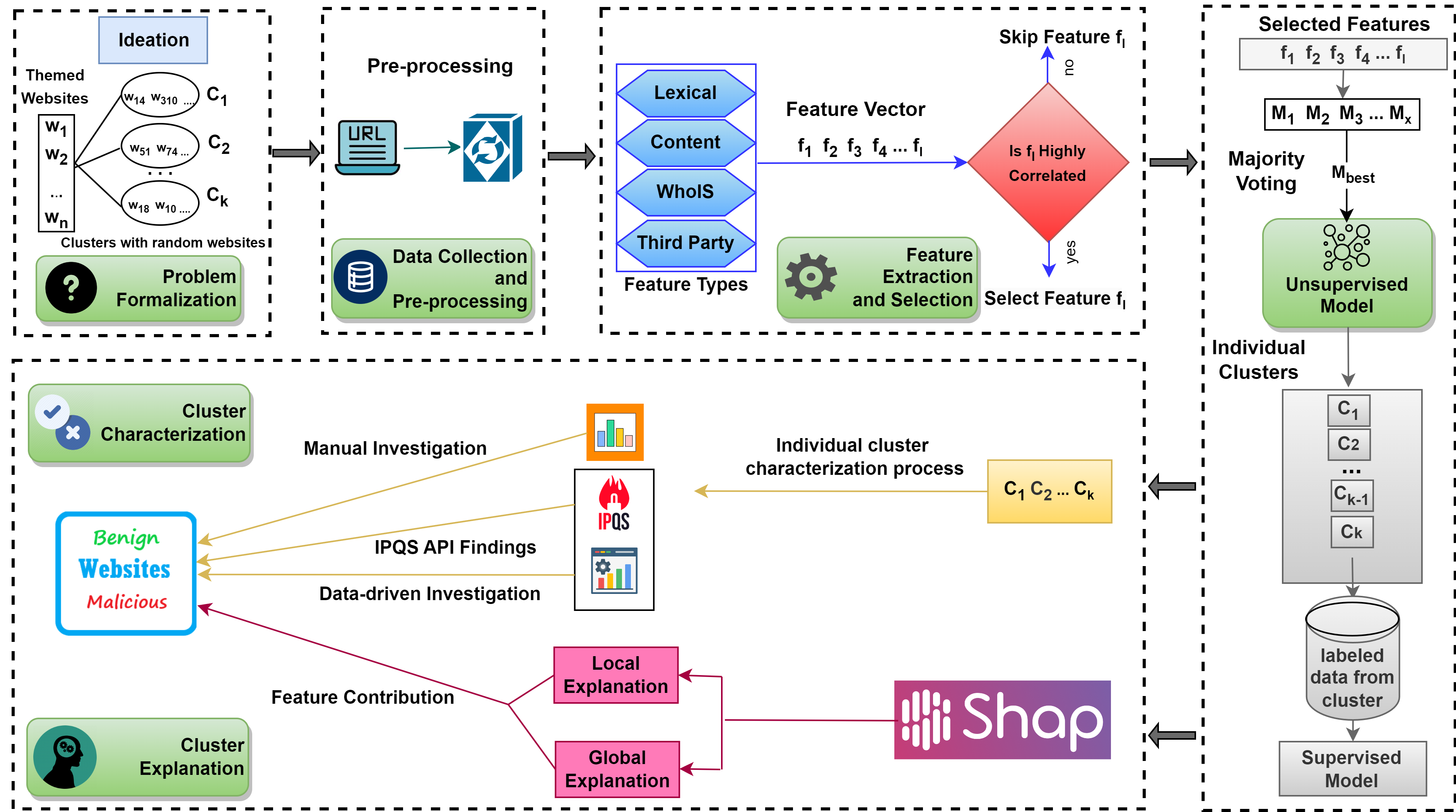}
    \caption{Overview of war-themed website campaigns' characterization methods}
\label{fig:overview_methods}
\end{figure*}

\vspace{-0.30em}
\subsection{Problem Formalization}
\label{sec:problem_formalization}
Consider a dataset of $n$ websites, denoted by
$W = \{w_1, w_2,\cdots, w_n\}$. To address the RQs, we formalize them as the following problem: Cluster the $n$ websites into $K$ groups denoted by $\{C_1, C_2, \cdots, C_K\}$, where $K<n$,
based on a set of $l$ features extracted from the website denoted by $f = \{f_1, f_2, \cdots, f_l\}$ and using $m$ different clustering models. We further set the cluster label for any individual website $w_i$ with a majority voting among models. To illustrate the voting, lets $C_k^{(j)}$ denote the $k$-th cluster produced by the $j$-th clustering model, where $k= 1, 2, 3, \cdots, K$ and $j=1,2,3,\cdots, m$. Let, $g_j: W \rightarrow \{1, 2, \cdots, K\}$ be the function representing the output of the $j$-th clustering model, assigning each website $w_i$ to one of the clusters. Next, we define a voting score $V_{i,k}$ for website $w_i$ to assign it to cluster $k$ based on our $m$ clustering models. The voting score for $w_i$ to belong to cluster $k$ is calculated as follows: 
\vspace{-0.5em}
\[ V_{i,k} = \sum_{i=1}^{m} \mathbb{I}(g_j(w_i) = k)
\]
Where $\mathbb{I}(.)$ is the indicator function-
\[
\mathbb{I}(g_j(w_i) = k) =
\begin{cases}
1 & \text{if } g_j(w_i) = k, \\
0 & \text{otherwise.}
\end{cases}
\]
Finally, the cluster assignment for website $w_i$ is determined as the function $h(w_i)$-
\[
h(w_i) = \arg\max_{k \in \{1, 2, \ldots, K\}} V_{i,k}
\]
The above method makes sure that the cluster with the highest total votes $V_{i,k}$ across the $m$ models is selected for $w_i$. For breaking a tie, a random selection of clusters can be applied. Additionally, we gather the clustering explanation results based on the feature set $f$, providing insights on general characteristics of any cluster $\forall_k C_k$ and why a certain website $w_i$ belongs to an individual cluster $C_k$. Our ultimate objective is to find if any cluster (or multiple clusters) $\exists_{k}C_k$ that can be labeled as potentially malicious (i.e., a high percentage of malicious sites). An overview of the proposed method is presented in Fig. \ref{fig:overview_methods}. 

\vspace{-0.35em}
\subsection{Data Collection and Pre-Processing}
During the relevant website data collection, we observed that war events, epidemics/pandemics, natural disaster events, or any calamities are correlated with the emergence of a high volume of event-themed website registrations in a shorter period. Thus, a dataset can be collected during the ongoing event by closely observing the domain registrations with specific keywords in the domain strings or through any third-party services that collect and share these types of domain name data, such as {\tt DomainTools.com}. However, one common issue is that the dataset is oftentimes \textit{unlabeled} with no ground truth. 
After having the data, we query all the websites to extract the relevant lexical, hosting infrastructure, and content features. Since only URL lexical features do not provide comprehensive characteristics and a convincing model performance, we try to analyze the websites' content data from the live websites during the analysis phase. 

\subsection{Feature Extraction and Selection} 
For clustering the event-themed websites, 
we define the following 34 features in
four categories: \textbf{Lexical}, \textbf{TLD}, \textbf{Content}, and \textbf{ WHOIS}.

\smallskip

\noindent\textbf{Lexical Features}. They are extracted from domain names only. It is important to mention that the keyword-based features and the breadth of the keywords are selected based on expert knowledge and deep manual investigation of these websites. We have defined the following list of lexical features--
\begin{itemize}
    \item \textbf{Domain Vowel Count ($f_1$)}: count of vowels (including `y') in a given domain. \{Returns: Count (INT)\}
    \item \textbf{Total Number of Domain Keywords ($f_2$)}: count of independent keywords in a domain name concerning a dictionary. \{Return: Count (INT)\}
    \item \textbf{Domain Unique Character Count ($f_3$)}: the number of unique characters in a given domain. \{Returns: Count (INT)\}.
    
    \item \textbf{Domain Letter Count ($f_4$)}: 
    count of letters (a-z or A-Z) in the domain name string. \{Returns: Count (INT)\}
    
    \item \textbf{Domain Digit Rate ($f_5$)}: the ratio of digit count and the total length of the domain, \{Return Type: Float, Returns: Score\}
    
    \item \textbf{Domain Hyphen Count ($f_6$)}: count for the number of hyphen (-) character in the domain name. \{Returns: Count (INT)\}
    
    \item \textbf{Presence of ``{\tt xn--}'' String ($f_{7}$)}: whether the domain name of a website starts with the string ``{\tt xn--}" (i.e., punycode). \{Returns: 1 (yes) $||$ 0 (no)\}
    
    
    \item \textbf{Presence of {\em fundraising} Related Keywords ($f_{8}$)}: 
    whether a domain name contains any money-related words from the following list [`money', `donate', `pay', `coin', `cash', `charity', `buy', `fund', `dollar'].
    \{Returns: 1 (yes) $||$ 0 (no)\}

    \item \textbf{Presence of {\em help} Related Keywords ($f_{9}$)}: 
    whether a domain name contains any of the help-related words from the following list [`help', `assist', `stand', `support', `aid', `save', `unite'].
   \{Returns: 1 (yes) $||$ 0 (no)\}

    \item \textbf{Presence of {\em cause campaign} Related Keywords ($f_{10}$)}: whether a domain name contains any cause campaign related word from the following list [`art', `concert', `pray', `care', `build', `hope', `food'].  \{Returns: 1 (yes) $||$ 0 (no)\}

    \item \textbf{Presence of {\em war} Related Keywords ($f_{11}$)}: whether a domain name contains any war and artillery related word from the following list [`war', `free', `fight', `tank', `weapon', `drone', `hero', `arm', `victory', `bullet', `gun'] \{Returns: 1 (yes) $||$ 0 (no)\}

    \item \textbf{Presence of `for' and `to' Keyword ($f_{12}$)}: whether a given domain contains `for' or `to' in it. \{Returns: 1 (yes) $||$ 0 (no)\}
    
    \item \textbf{Number `4' and `2' Instead of Word ($f_{13}$)}: whether digit `4' or `2' is present in a given domain name, where `4' and `2' represents word `for' and `to', respectively. \{Returns: 1 (yes) $||$ 0 (no)\}
    
    \end{itemize}

\noindent\textbf{TLD-based Features}. Features are extracted from the TLD of domain names:
\begin{itemize}
    \item \textbf{Is Cheap TLD ($f_{14}$)}: whether the TLD of the domain name is from our curated list of 77 cheapest TLDs (less than \$2 per year registration fee) during the year 2022. \{Returns: 1 (yes) $||$ 0 (no)\}
    
    \item \textbf{Is Multiple TLD Combined ($f_{15}$)}: whether the TLD in the domain contains country or other TLDs. \{Returns: 1 (yes) $||$ 0 (no)\}
    
    \item \textbf{Spamhaus TLD Badness Score ($f_{16}$)}: Spamhaus \cite{Spamhaus} scoring produces \textit{Badness Score} for a given TLD based on the number of websites seen with a particular TLD and the number of websites flagged as malicious. A higher score indicates a bad TLD. For this feature, we consider the badness score from December 2023 from the Spamhaus websites. \{Returns: Score (Float)\}

\end{itemize}

\noindent\textbf{Content Features}. 
These features are extracted from website contents (e.g., SSL Certificate), mainly querying the results from the final landing page. Again, it is important to highlight that we have relied on domain knowledge and manual investigation to incorporate the presence of certain keywords and token-based features. The final list of features includes:
    
\begin{itemize}
    \item \textbf{Is SSL Certificate Retrievable  ($f_{17}$)}: whether the SSL certificate associated with a website is retrievable or not. 
    \{Returns: 1 (yes) $||$ 0 (no)\} 
 
    \item \textbf{Final Landing Page Size  ($f_{18}$)}: The overall byte size of the raw contents of the final landing page of the website in kilobytes. \{Returns: Score (Float)\}

    \item \textbf{Has Same Final Landing Domain  ($f_{19}$)}: whether the final landing page is the same as the queried domain (redirection). 
    \{Returns: 1 (yes) $||$ 0 (no)\}

    \item \textbf{Unique Internal Website Page Count ($f_{20}$)}: total number of unique page links in the landing page associated with the queried domain. \{Returns: Count (INT)\}

    \item \textbf{Unique External Website Link Count ($f_{21}$)}: total number of unique page links that have a different domain than the queried one found in the final landing page. \{Returns: Count (INT)\}

    \item \textbf{Avg. Length of the External Websites' Link ($f_{22}$)}: average length of all unique external websites' links found in the landing page. \{Returns: Score (Float)\}

    \item \textbf{Std. of Length of the External Websites' Link ($f_{23}$)}: standard deviation of length of all unique external websites' links found in the final landing page. \{Returns: Score (Float)\}

    \item \textbf{Ratio of Shortened URL in External Website Links ($f_{24}$)}: ratio of shortened URL (checked with $1,542$ URL shortener) within all unique external websites' links found in the final landing page. \{Returns: Score (Float)\}

    \item \textbf{Ratio of Social Media URL in External Website Links ($f_{25}$)}: ratio of some predefined social media links (Facebook, Linkedin, Twitter, Instagram, Whatsapp, Snapchat, Youtube, Telegram) within the all unique external websites' links found in the final landing page. \{Returns: Score (Float)\}

    \item \textbf{Has External Form Link  ($f_{26}$)}: whether the final landing page has external form links (i.e. Google, Firebase forms). \{Returns: 1 (yes) $||$ 0 (no)\}

    \item \textbf{Has PayPal Payment Link  ($f_{27}$)}: whether the final landing page has an external PayPal link for payment. \{Returns: 1 (yes) $||$ 0 (no)\}

    \item \textbf{Has Crypto Related Word Token  ($f_{28}$)}: whether the final landing page contains some predefined Crypto-related keywords from the following list [`crypto', `block', `chain', `blockchain', `wallet', `nft',  `coin', `thereum', `bitcoin', `altcoin', `btcc', `trx', `doge', `solana', `cardano', `usdc', `xrp', `tether', `usdt', `eth', `sol', `btc']. \{Returns: 1 (yes) $||$ 0 (no)\}

    \item \textbf{Has Bank Related Word Token  ($f_{29}$)}: whether the final landing page contains some predefined Bank related keywords from the following list [`bank', `banking', `monobank', `account', `acc', `accnt', `branch', `routing', `rout']. \{Returns: 1 (yes) $||$ 0 (no)\}

    \item \textbf{Has Card Related Word Token  ($f_{30}$)}: whether the final landing page contains some predefined Card-related keywords from the following list [`creditcard', `credit', `card', `mastercard', `debit', `visa', `cards', `kreditkarte', `kredit', `karte']. \{Returns: 1 (yes) $||$ 0 (no)\}

\end{itemize}

\noindent\textbf{WHOIS Features}. Features are related to the 
WHOIS data of a website:
    
    \begin{itemize}
     \item \textbf{Is WHOIS Registrar available  ($f_{31}$)}: whether the WHOIS data of a website is available or not. \{Returns: 1 (yes) $||$ 0 (no)\}
    \item \textbf{Is Name Server available  ($f_{32}$)}: whether the name server from the WHOIS data is available or not for a given website. \{Returns: 1 (yes) $||$ 0 (no)\}
    \item \textbf{Is Self-Resolving Name Server ($f_{33}$)}: whether the name server from WHOIS data is self-resolving or not. 
    \{Returns: 1 (yes) $||$ 0 (no)\}
    \item \textbf{Total Number of Name Servers ($f_{34}$)}: the total number of name servers from WHOIS data for a given website. \{Returns: Count (INT)\}.
    
\end{itemize} 


Although keyword-related characteristics, specifically $f_{28}$, $f_{29}$, and $f_{30}$, are primarily extracted from English textual content, their design allows them to function as language-agnostic characteristics, extending their applicability in other languages.

\noindent \textbf{Feature Selection.} To select the best set of effective characteristics, we have adopted a pairwise Pearson correlation-based approach \cite{benesty2009pearson} to remove some highly correlated elements such that 
if $\forall_{i,j}Corr(f_i, f_j) > Th_{corr}$, then we remove either feature $f_i$ or feature $f_j$ from the final selected feature set. 

\vspace{-0.45em}
\subsection{Unsupervised Machine Learning: Clustering}
\label{sec:clustering_model}
The event-themed website dataset $W$ is unlabeled, as very few malicious activities are caught at the beginning of an ongoing event. Hence, we need to incorporate an unsupervised machine learning approach (i.e., clustering models) to group similar websites and find common characteristics between these groups of websites. Hence, in this module, we choose $m$ number of state-of-the-art unsupervised clustering models denoted as $M_1, M_2, \cdots, M_m$ such as K-means \cite{IKOTUN2023178}, K-medoids \cite{kmedoid_park2009simple}, Hierarchical clustering \cite{nielsen2016hierarchical}, \ignore{\color{red}DBSCAN (Density-Based Spatial Clustering of Applications with Noise) \cite{schubert2017dbscan}} or GMM Clustering (Gaussian Mixture Model) \cite{weber2022gaussian} to find $k$ different website clusters where any $k$-th cluster $C_k$ contains a list of unique websites from the set of websites $W$. Formally, $|C_1| + |C_2| + \cdots + |C_k| = |W|$. Depending on the chosen clustering model, in some models, $k$ must be predefined for that clustering algorithm. 
 To determine the value of $k$ in each selected clustering model, we adopted the Silhouette score \cite{Shillouette,silhouttee_just_1_januzaj2023determining} as a baseline to reach a consensus among models. This score is a measure of the similarity of an object to its cluster while taking into consideration its similarity to the other clusters. This score is within the range [-1,1], 
 where a high value indicates that the object is well-matched to its own cluster and poorly matched to the other clusters. We have also used other standard evaluation techniques (i.e., elbow graphs, dendrogram) as a secondary measure to show the effectiveness of the Silhouette score-based approach.
After generating the clusters with the selected $m$ models, we proceed to find the most effective way of grouping the websites. To achieve that, we propose to incorporate a majority voting scheme among the $m$ clustering models. 
For this approach to work, the number of clusters ($k$) should be equal. 

 


\vspace{-0.45em}
\subsection{Feature-Based Cluster Explanation} 
In this module, we get more insights into individual clusters with feature-based explanations. We use the popular SHAP (\textbf{SH}apley \textbf{A}dditive ex\textbf{P}lanation) \cite{lundberg2017unified} module in XAI (explainable AI) for unsupervised models \cite{unservised_explanation}. Once we find the corresponding clusters for each website $w_i \in W$, we consider each cluster $\forall_{k}C_k$ as a class label for extracting the explanation of that cluster using a supervised ML model. 
Hence, for each $k$th cluster, $C_k$, we generate a global summary plot with respective features to understand the top contributing features and their 
ranges of values. 

\ignore{
\begin{figure}
    \centering
    \includegraphics[width=1\textwidth]{figures/SHAP.png}
    \caption{Explanation for the unsupervised clustering using SHAP}
    \label{fig: Explanation for the clusters using SHAP}
\end{figure}

}

\vspace{-0.45em}
\subsection{Data-Driven Cluster Characterization}
In this module, we find characteristics of certain clusters based on the deep dive data analysis of features, the website contents, and structures. 
We also analyze to find any potential malicious cluster (or clusters) that accommodate high percentages of malicious websites. This deep-dive analysis further helps analysts to shed light on an important aspect of course-of-action decision-making on websites such as {\tt take-down} or {\tt notify owner} \cite{pritom_law_support_cns22}. 
For example, we want to investigate why any cluster $C_i$ is different than another cluster $C_j$ based on the corresponding websites' features and their structural differences. Moreover, after characterization, we incorporate a third-party threat intelligence service like IPQS (IP Quality Score) \cite{IPQualityScore} or {\em VirusTotal.com}, to assess and validate our clustering outcomes. As in the case of retrospective studies, some websites may not be available in live status during our analysis, 
the defenders may consider a subset of live-themed websites, $W_{live} \subseteq W$. These third-party intel can help to analyze and validate if certain clusters can be deemed as malicious than other clusters based on the high percentage of malicious websites within a cluster. However, these tools are not completely reliable when an event is ongoing during real-time analysis. Thus, this tool cannot be used as a method of labeling or generating ground truth in the early phase of ongoing event-themed website campaigns. 

%% file: studyresults.tex

\subsection{Data Collection and Pre-Processing}
The analysis pertaining to websites themed around the Russia-Ukraine war was initiated by the acquisition of a dataset sourced from {\tt DomainTools.com}, obtained in its original state without any labels. The parameters guiding our data collection procedure entailed the retrieval of domain names associated with websites registered between February 25, 2022, and July 25, 2022, (The most recent Russia-Ukraine conflict began on February 24, 2022), containing either the terms ``Ukraine" or ``Ukrainian" within their domain name.


\begin{figure}[!htbp]
    \centering
\includegraphics[width=0.75\columnwidth]{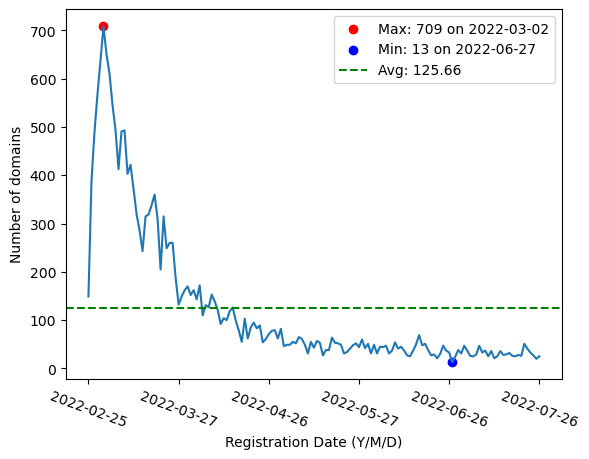}
\caption{Domain Registration trends for websites from Domaintools}
\vspace{-0.5em}
\label{fig:dom_name_period}
\end{figure}

Our data comprises 151 consecutive days on which domains are reported in DomainTools. Figure \ref{fig:dom_name_period} denotes the daily frequency of websites collected from DomainTools. 
The declining trends in website registration after the first month of the event start date agree with other reported threats related to events, such as COVID-19 themed websites \cite{lallie2020cyber_COVID19Paper,COVID19_data_checkphish,mir_covid_website20}. As per our observation for this case study, we can consider the first 7-10 days as the `early-stage', as more than a thousand websites were registered within the first week of the event start date. 
Following the compilation of the dataset, we have merged data from different days into one master list and removed any duplicates to generate themed-website dataset $W$, where $|W| = 18,321$ websites on Q4 2023. Since we are retrospectively conducting this study, we want to get only the live websites during our feature extraction process (Q2 2024), resulting in $|W_{live}| = 2,118$ websites for this case study. We provide further details of this case study, experiments, and results in a GitHub repository located at {https://github.com/MarazMia/Theme-Threat-Research-101}.

\subsection{Feature Extraction and Selection}
We have collected all 34 features, namely $\{f_1, f_2, \cdots, f_{34}\}$ 
and after requiring those features, we select a subset from the above 34 features to remove some of the pair-wise highly correlated ones that can lead to the multicollinearity issue. 
We generate a Pearson feature correlation matrix ($34 \times 34$) to observe and remove some features that have a correlation coefficient higher than 0.6 ($Th_{corr}=0.6$). Finally, 28 features have been finalized after removing the following features from the feature list: $f_1$, $f_2$, $f_3$, $f_{32}$, $f_{33}$, and $f_{34}$. 

\subsection{Unsupervised Clustering}
 
In this case study, we have chosen four unsupervised clustering algorithms: K-Means, K-Medoids, Hierarchical, and GMM. To find the cluster size for each approach, we consider two justifications: (i) baseline criteria from {\em Silhouette} scores and (ii) cross-checked with other standard methods (elbow graphs, dendrogram). It is important to note that the Silhouette score-based cluster number selection can be automated. 

\begin{figure}[!t]
    \centering
    \begin{subfigure}[b]{0.74\columnwidth}
        \centering
        \includegraphics[width=\textwidth]{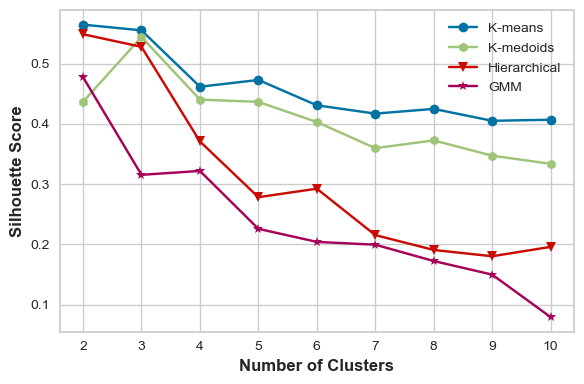}
        \caption{Average Silhouette score of clusters}
        \label{fig:avg_silhouette}
    \end{subfigure}
    \hfill
    \begin{subfigure}[b]{0.74\columnwidth}
        \centering
        \includegraphics[width=\textwidth]{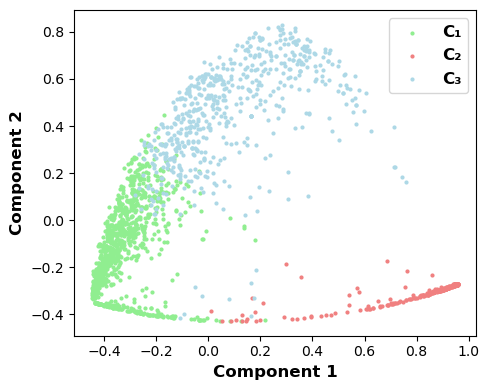}
        \caption{PCA plots based on 3 clusters}
        \label{fig:pca}
    \end{subfigure}
    \caption{Visualization of clustering results: (a) Average Silhouette score and (b) PCA plot for 3 clusters}
    \label{fig:clustering_results}
\end{figure}

\noindent \textbf{(a) K-Means} is an average distance or centroid-based model that provides strong insights into the location of each cluster. However, the cluster size $k$ must be declared at the beginning of this model. From our baseline estimation (as described in section 3.4), 
we get the value of $k = 3$ (blue curve in fig \ref{fig:avg_silhouette}). Further, we observe the Elbow graph for the  Calinski-Harabasz score \cite{calinski1974dendrite} and find the elbow point at $k=3$, which resonates with our baseline approach (Figure A.1 in Github Readme). 

\noindent \textbf{(b) K-Medoids} is a very similar approach to K-Means except for the median value for the centroid. This method is robust against the noise and outliers in the data. Like as K-means, we need to specify the number of cluster $k$ and we again use the same primary evaluation technique of Silhouette score (green curve in fig \ref{fig:avg_silhouette}) and then generate the elbow graph (Figure A.1 in Github Readme) to discover $k=3$ in both cases.

\noindent \textbf{(c) Hierarchical clustering} with Gower distance is selected in this study since we deal with both binary and continuous feature data \cite{nielsen2016hierarchical}. In this clustering approach, we do not need to specify the value of $k$ in the first place. The $k$ can be observed from the Dendrogram plots (Figure A.2 in Github Readme), which shows that $k=3$ is a good cutoff point and agrees with the baseline method's findings (red curve in Figure \ref{fig:avg_silhouette}).  

\noindent \textbf{(d) GMM} is another sophisticated clustering approach that models data as a mixture of Gaussian distributions \cite{weber2022gaussian}. This method is also helpful in cases of overlapping clusters. It provides a probability for each data point belonging to each cluster, which can be valuable for softer clustering tasks where uncertainty is present. Again, to determine the value of $k$, we first use our baseline estimation, resulting in $k=2$, as shown in Figure \ref{fig:avg_silhouette}-- pink curve. Later, we use the Akaike Information Criterion (AIC) and Bayesian Information Criterion (BIC) scores \cite{akaike2011akaike} with the elbow graph, which shows that $k=3$ as the number of clusters. The results are presented with details in the GitHub Readme file \textit{section A.3} due to space constraints. 

\ignore{
\begin{figure}[!t]
    \centering
\includegraphics[width=.90\columnwidth]{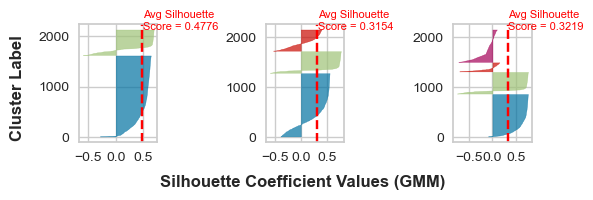}
    \caption{GMM Cluster's average Silhouette score for $k=2, 3, 4$. All the criteria are fulfilled when $k=3$}
    \label{fig:individual_silhouette}
\end{figure}
}

\begin{table}[!h]
\caption{Website Distribution Among Clusters (All Models)}
\centering
\resizebox{0.65\columnwidth}{!}{
\begin{tabular}{|p{2cm}|*{3}{p{0.9cm}|}}
\hline
\multirow{2}{*}{\textbf{Clustering 
Model}} & \multicolumn{3}{c|}{\textbf{Website Instance Counts}} \\
\cline{2-4}
& $C_1$ & $C_2$ & $C_3$\\
\hline
K-Means & 1185 & 379 & 554 \\
\hline
K-Medoids & 1085 & 385 & 648\\
\hline

Hierarchical & 1095 & 412 & 611\\
\hline


GMM & 1263 & 431 & 424\\
\hline
Majority Voting & 1149 & 388 & 581\\
\hline

\end{tabular}
}
\label{table:Cluster_Results}
\end{table}
For all 4 selected clustering models, it is evident that a value of $k=3$ can be justified as the final number of clusters based on majority agreement. 
Next, Table \ref{table:Cluster_Results} provides the distributions of websites among the clusters for all the selected models and the final distribution after the majority voting, which shows $|C_1| = 1,149$ websites, $|C_2| =388$ websites, and $|C_3| = 581$ websites after the voting. 
We hereby address the \textbf{RQ1}. From the Principal Component Analysis (PCA) plot in Figure \ref{fig:pca}, we can also observe that cluster $C_2$ is well separated compared with the other two clusters $C_1$ and $C_3$.

\subsection{Clustering Explanation With SHAP}
One of our important objectives is to know why a particular group of websites is forming a specific cluster. To address the feature-based explanation of clusters, we first convert it into a multi-class classification where classes are the corresponding cluster labels. 
In our case, we have 3 clusters, meaning 3 different classes for training and testing. We leverage {\em XGBoost} supervised classifier and later added a feature-based SHAP explanation for each cluster, as XGBoost integrates well with SHAP and outperforms other ML models. Finally, we split the data into a $80:20$ train-test set. The hyper-parameters are set as {\tt max\_depth=2}, {\tt random\_state=40}, {\tt n\_estimator=200}, {\tt learning\_rate=0.25}, {\tt min\_child\_weight = 3}, and {\tt gamma=0.5}. 

\ignore{
\begin{table}[!b]
\caption{Hyperparameters Used in XGBoost Model}
\centering
\resizebox{0.38\textwidth}{!}{
\begin{tabular}{|*{4}{c|}}
\hline
\textbf{Parameter} & {\textbf{Value}} & \textbf{Parameter} & {\textbf{Value}}\\
\hline
objective & multi:softprob & max\_depth & 2\\
\hline
random\_state & 40 & n\_estimators & 200 \\
\hline
learning\_rate & 0.25 & min\_child\_weight & 3 \\
\hline
gamma & 0.05 & colsample\_bytree & 0.1 \\
\hline

\end{tabular}
}
\label{tab:xgb_param}
\end{table}
}

\begin{figure}[!t]
    \centering
    \begin{subfigure}[b]{0.85\columnwidth}
        \centering
        \includegraphics[width=\textwidth, height=.75\columnwidth]{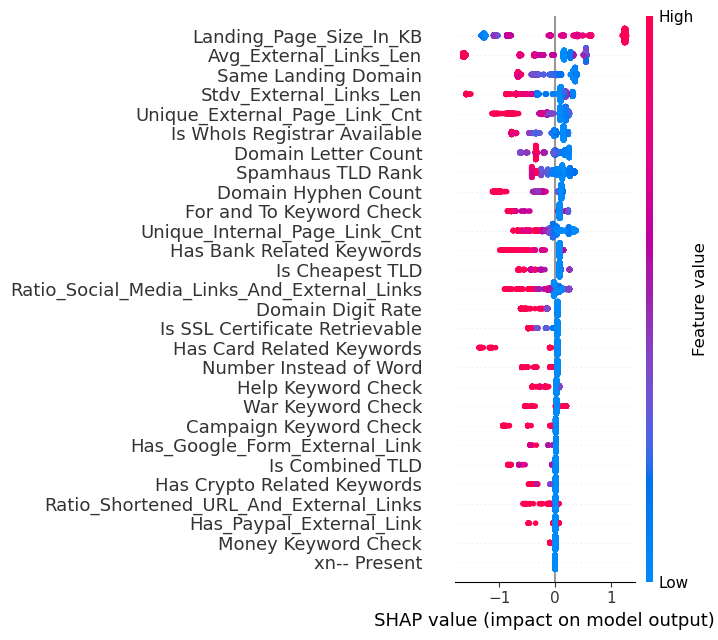}
        \caption{Summary plot for cluster $C_1$}
        \label{fig:exp_c1}
    \end{subfigure}
    \hfill 
    \begin{subfigure}[b]{0.85\columnwidth}
        \centering
        \includegraphics[width=\textwidth, height=.75\columnwidth]{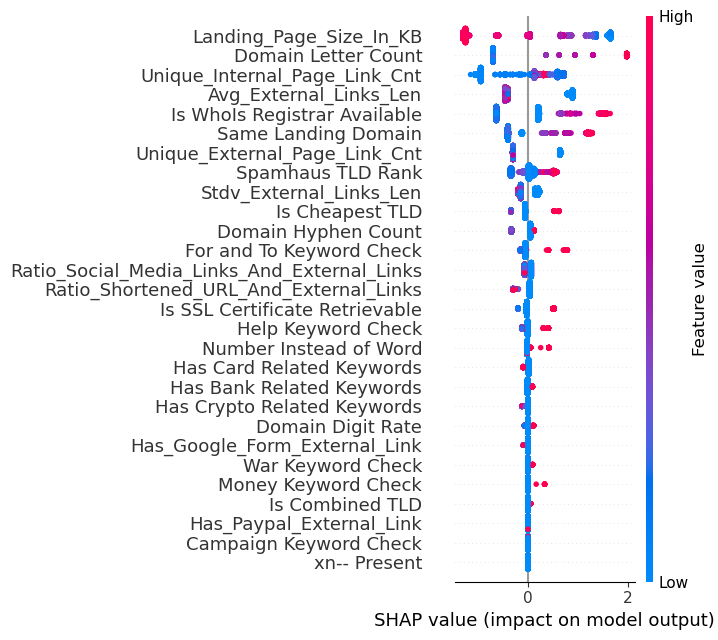}
        \caption{Summary plot for cluster $C_2$}
        \label{fig:exp_c2}
    \end{subfigure}
    \begin{subfigure}[b]{0.85\columnwidth}
        \centering
        \includegraphics[width=\textwidth, height=.80\columnwidth]{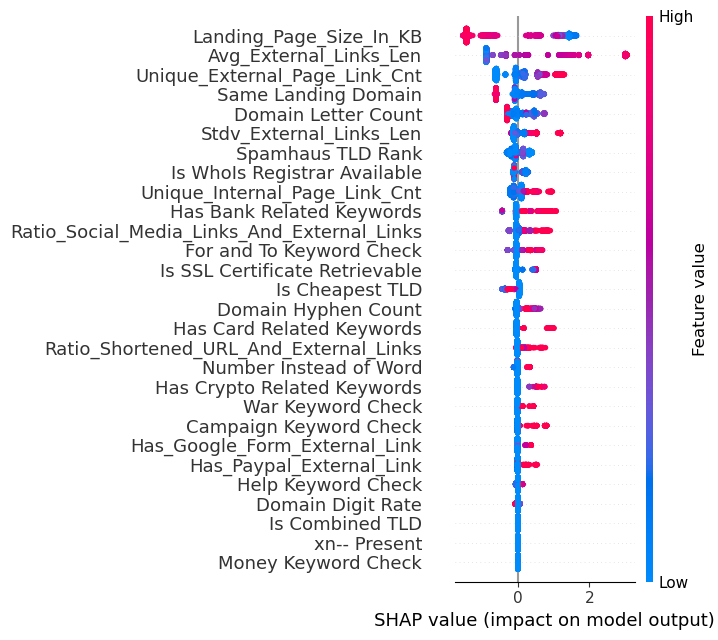}
        \caption{Summary plot for cluster $C_3$}
        \label{fig:exp_c3}
    \end{subfigure}
    \caption{SHAP summary plots for clusters $C_1$, $C_2$, and $C_3$}
    \label{fig:exp_clusters}
\end{figure}

Next, we observe the XGBoost model's evaluation metrics such as-- {\em accuracy}, {\em precision}, {\em recall}, and {\em F1-score} which are all equal to $0.99$ (i.e., indicating the model is accurate) on the test data (Confusion Matrix Heatmap C.1 in Github Readme). 
This verification step for using the supervised ML model is crucial for enhancing the transparency and trust in the clustering. Next, for the individual clusters ($C_1$, $C_2$ and $C_3$), we further observe the feature explanations through the summary plots as presented in Figure \ref{fig:exp_c1}, Figure \ref{fig:exp_c2}, and Figure \ref{fig:exp_c3}, respectively. A SHAP summary plot shows feature importance and its effects on model predictions. Each point represents a SHAP value for a feature and an instance, with color indicating the feature's value (e.g., red for high and blue for low). The $x$-axis shows SHAP values, where positive values push predictions higher and negative values push them lower for that class. Features are ordered by importance, with larger spreads indicating greater variability in their impact. This plot not only highlights which features are most influential but also shows how different values of each feature affect predictions, providing insight into the model’s behavior.

According to Figure \ref{fig:exp_c1} (summary plot for cluster $C_1$), websites in the $C_1$ cluster have a uniform distribution of lower feature's value except for the {\em landing page size}, {\em presence of for\_to} and {\em war-related keywords} features. From the top-most important features, {\em landing page size}, {\em external link length}, {\em same landing domain}, {\em is WHOIS registrar available}, and {\em domain letter counts} are notable. From Figure \ref{fig:exp_c2} (summary plot for cluster $C_2$), features as {\em domain letter counts}, {\em is WHOIS registrar available}, {\em same landing domain}, {\em Spamhause TLD rank}, {\em cheapest TLD}, {\em presence of for\_to}, {\em presence of help related keywords in domain} and {\em presence of fundraising related keywords in domain} are displaying a higher value range while {\em landing page size}, {\em internal page link count}, and {\em external link length} has lower feature values in contrast. Lastly, from Figure \ref{fig:exp_c3} (summary plot for cluster $C_3$), we observe a more versatile feature distribution of higher values for the {\em external link length}, {\em unique external page count}, {\em internal page count} features while showing lower feature values for {\em landing page size}, {\em is same landing page}, {\em domain letter count}, {\em Spamhaus TLD rank}, {\em is WHOIS registrar available} and {\em is SSL retrievable}. In terms of feature SHAP values, for cluster $C_2$ and cluster $C_3$, we see a more dispersed distribution of SHAP values mostly closer to the positive direction while in cluster $C_1$, these values are centered around 0. 
To grasp the whole picture, we provide some further analysis with other SHAP plots in the project's GitHub repository Readme \textit{section C}.

\begin{table}[!htbp]
\centering
\caption{Data-driven Insights For Each Cluster Based On Feature Attributes}
\resizebox{0.95\columnwidth}{!}{
\begin{tabular}{|p{0.75cm}|*{2}{p{4cm}|}}
\hline
\multirow{2}{*}{\textbf{Cluster}} & \multicolumn{2}{|c|}
{\textbf{Characteristics}} \\
\cline{2-3}

 \textbf{(size)}& \textbf{Based on Categorical Features} & \textbf{Based on Numerical Features}\\
\hline
 
$C_1$ (1,149) & highest landing page size (avg. 263.71 KB), only 4.43\% SSL retrievable, less usage of {\em war} keywords (only 4.1\% of websites), lowest usage of cheap TLD (14.71\%) & 
lowest Spamhaus avg. TLD badness score of 0.11, higher cases of social media links in external links (avg. ratio is 0.32)\\
\hline
$C_2$ (388) & lowest landing page size (avg. 2.57 KB), less usage of {\em bank} keywords (1.03\% websites), highest SSL certificate retrievable (31.70\%) & no cases of external links in home page(0\%), higher cases of hyphened domain (avg. ratio of 0.35)\\
\hline
$C_3$ (581) & higher cases of bank-related keywords (13.61\%) and for\_to keywords (17.51\%) & highest external links length (avg. 55.97), higher cases of hyphened domain (avg. ratio of 0.35)\\
\hline
\end{tabular}
}
\label{table:cluster attributes}
\end{table}

\subsection{Cluster Characterization}

Our ultimate objective is to identify potential cluster (or clusters) that show highly malicious characteristics. 
To achieve that and address \textbf{RQ2}, we have to rely on the following analysis-- (i) feature-based; (ii) third-party intelligence (IPQS and VT) based; (iii) manual analysis on selected samples; and (iv) website contents and structure-based.

\subsubsection{Feature and Attribute-based Analysis.} Although we already got some of the information from the previous section's SHAP-based analysis on the features' contribution and value ranges for the clusters, we want to explore more to make sure that we do not miss any important feature pattern. After the investigation, we have identified several key characteristics of each cluster as presented in Table \ref{table:cluster attributes} highlighting insights based on the categorical and numerical features that are mostly missing in the previous SHAP-based analysis. 

\begin{table*}[!htbp]
    \centering
    \caption{Queried IPQS and VirusTotal Flags Counts for Three Clusters}
    \resizebox{0.85\textwidth}{!}{
    \begin{tabular}{|c|c|c|c|c|c|c|c|c|c|}
        \hline
        \multirow{2}{*}{\textbf{Cluster Id}} & \multicolumn{5}{c|}{\textbf{IPQS Flag}} & \multicolumn{4}{c|}{\textbf{VT Flag}} \\
        \cline{2-10}
        & Spamming  & Malware & Phishing & Suspicious &  Risk Score & Malicious & Suspicious & Clean & Not Found \\
        \hline
        $C_1$ & 7 (0.61\%) & 2 (0.17\%) & 4 (0.35\%) & 522 (45.4\%) & 154 (13.4\%) & 34 (2.96\%) & 19 (1.65\%) & 399 (34.7\%) & 772 (67.2\%) \\
        \hline
        $C_2$ & 5 (1.29\%) & 0 (0\%) & 0 (0\%) & 307 (79.1\%) & 267 (68.8\%) & 13 (3.35\%) & 5 (1.29\%) & 99 (25.5\%) & 250 (64.4\%) \\
        \hline
        $C_3$ & 5 (0.86\%) & 3 (0.52\%) & 4 (0.69\%) & 261 (44.9\%) & 118 (20.3\%) & 20 (3.44\%) & 2 (0.34\%)  & 178 (30.6\%) & 392 (67.5\%) \\
        \hline
    \end{tabular}}
    \label{tab:ipqs_flag_cnt}
\end{table*}

\begin{table}[!htbp]
\caption{Distribution of IPQS Malicious Flagged Websites Among Clusters}
\centering
\resizebox{0.95\columnwidth}{!}{
\begin{tabular}{|p{2cm}|*{3}{p{1.90cm}|}}
\hline
\multirow{2}{*}{\textbf{Clustering 
Model}} & \multicolumn{3}{l|}{\textbf{IPQS Malicious Count (Percentage)}} \\
\cline{2-4}
& $C_1$ & $C_2$ & $C_3$\\
\hline
K-Means & 137 (11.56\%) & 261 (68.87\%) & 112 (20.22\%) \\
\hline
K-Medoids & 119 (10.97\%) & 267 (69.35\%) & 124 (19.14\%)\\
\hline

Hierarchical & 127 (11.6\%) & 264 (64.08\%) & 119 (19.48\%)\\
\hline


GMM & 143 (11.32\%) & 267 (61.95\%) & 100 (23.58\%)\\
\hline

Majority Voting & 133 (11.58\%) & 262 (67.53\%) & 115 (19.79\%)\\
\hline

\end{tabular}
}

\ignore{
\centering
\resizebox{0.48\textwidth}{!}{
\begin{tabular}{|p{2.35cm}|*{3}{p{1.525cm}|}}

\end{tabular}
}
}
\label{table:ipqs_Cluster_Results}
\end{table}


\subsubsection{Third-Party Intelligence-based Analysis.}
We have queried both IPQS and VirusTotal (VT) API services as our third-party intelligence sources, but eventually selected IPQS because from the VT query results, most of the websites are not previously observed or reported by the vendors evidenced by the `Not Found' status reported in Table \ref{tab:ipqs_flag_cnt}.
IPQS provides the following binary flags on {\em Spamming}, {\em Malware}, {\em Phishing}, and {\em Suspicious} labels along with an integer {\em risk-score} between 0 to 100. For this experiment, we use IPQS query results from the Q4 of 2023 with the following condition -- If at least one of the boolean flags ({\em Spamming}, {\em Malware}, {\em Phishing}, {\em Suspicious}) is true and ({\em risk-score}$\geq 75$), then we consider the website as {\em malicious}; otherwise as {\em benign}. IPQS also provides other indicators like {\em Parking}, {\em adult}, and {\em unsafe}, which are overlapping with the above four selected flags.

Table \ref{table:ipqs_Cluster_Results}  shows that 67.53\% of websites of the $C_2$ are flagged as malicious by IPQS. On the other hand, cluster $C_1$ has around 11.58\% malicious websites (lowest among the clusters) while cluster $C_3$ has 19.79\% malicious websites according to IPQS. 
Although IPQS is a good pre-existing tool for this kind of analysis, it can not be trusted completely, since throughout our investigation, we have found evidence where IPQS failed to identify or flag some truly malicious websites that appeared malicious through our manual analysis, meaning the above percentages are likely conservative estimates of maliciousness. Moreover, IPQS's effectiveness on newer websites is still unknown. Another drawback of IPQS is that, most of the websites flagged by IPQS have the {\em Suspicious} flag being true as provided in Table \ref{tab:ipqs_flag_cnt} while the other flags such as {\em Spamming}, {\em Phishing}, and {\em Malware} have low numbers that fails to provide any insights on why exactly those websites are flagged as suspicious and what are the associated threats.

\subsubsection{Manual Analysis of Selected Sample Websites.}
As we do not have any trusted ground-truth labels for the websites, it's very cumbersome to provide the cluster purity measurement. Thus, to further evaluate the clustering approach, we have randomly selected $150$ websites ($50$ from each cluster) and manually analyzed these websites for their characteristics and behaviors. However, as there exists a time gap between the feature extraction and the manual investigation stage, some content of the website may have been altered, which we could not keep track of. In summary, we have checked the websites to find if any of the following threat indicators exist such as -- (a) malicious activities (phishing, browser's security message, malicious payloads and links), (b) fake donation, charity or other campaigns and crypto payment scams, (c) redirected website and links that lead to other shady websites, (d) parked domains for future attacks, (e) under-construction websites and (f) compromised websites (blank page, irrelevant contents, erroneous live websites). From our analysis, we found the following interesting facts about the clusters--

\begin{figure}[!htbp]
    \centering

\includegraphics[width=0.95\columnwidth]
{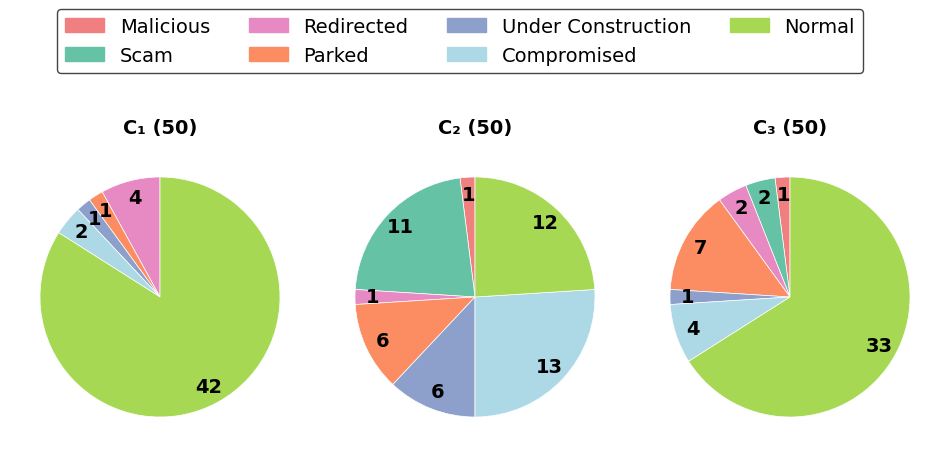}
    \caption{Manual analysis of 50 websites from each cluster}
    \label{fig:man_analysis}
\end{figure}

\textbf{Cluster $C_1$:} Out of 50 websites, most are online fundraising and donation groups with vetted organizational banks and PayPal accounts. These websites also include blogs, e-commerce sites, game sites, and volunteer with the content appearing to be well aligned and meticulously crafted. We have rarely found any examples of parked domains and very few cases of crypto donations. Nevertheless, we have discovered 4 redirected, 1 parked, 1 under construction, and 2 compromised websites, which account for 16\% of the manually investigated websites from this cluster (provided in diagram \ref{fig:man_analysis}). 

\textbf{Cluster $C_2$:} Among the 50 selected websites, we have observed 38 websites (76\%) being suspicious, of which 11 fall in the scam category. The scam websites can be further divided into 5 sub-categories: 
\begin{itemize}
    \item Donation scam (i.e., financial transaction): 3 sites
    \item Charity scam (i.e., financial transaction + other forms of charitable items): 2 sites
    \item Crypto scam: 3 sites
    \item Publicity scam (i.e., leveraging event for personal interest): 2 sites
    \item Service scam (i.e., job hiring, concerts, art showcase): 1 job hiring site
\end{itemize}
We have also discovered that in most cases, the crypto wallet and bank accounts (mostly personal) are provided in the raw text. Additionally, the payment links are not posted on the home page and would redirect 
to another domain than the original one. 
To make the donation campaign more appealing to the viewers, some websites have displayed the amount being collected so far from the ongoing donation campaign. Moreover, we have observed similar interfaces being used on multiple websites, meaning the same attacker groups might have weaponized multiple domains with the same content to scale the attacks. Another interesting find is that 3 particular websites were redirected to the same landing website that checks for human confirmation against DoS attack (e.g., \textit{Press ``Allow'' to prove you are not a robot}) and after the check, there was no action, which led us to label these websites as compromised. In one particular case, we find a website that provides a QR code for a malicious Android application download access that is no longer available in Google Play. Furthermore, out of the $13$ compromised sites, $2$ had \textit{backend directory} shown on the landing page.

\ignore{
\begin{figure}[!b]
    \centering
\includegraphics[width=0.50\linewidth]
{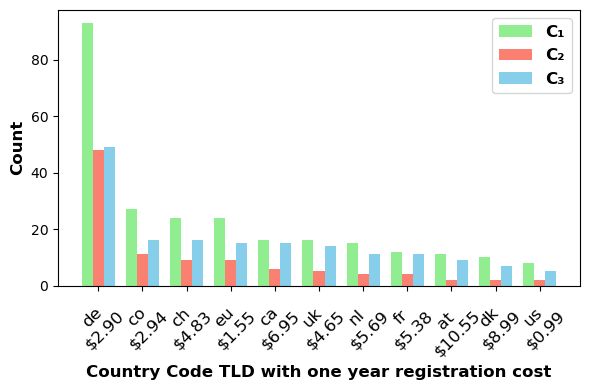}
    \caption{Analysis of top 11 ccTLDs}
    \label{fig:cc_tld}
\end{figure}
}

\textbf{Cluster $C_3$:} Most websites in this cluster show similar characteristics as in cluster $C_1$. Also, we encounter the frequent use of external social media links within these websites, with occasional cases of parked domains. Another interesting finding is that we have seen a higher number of non-English language (mostly Ukrainian) websites in this cluster compared with the other clusters. Figure \ref{fig:man_analysis} also shows that we encounter only 1 malicious website flagged by the browser as `\textit{Not Secure}'. 
Additionally, we have found 2 scamming, 7 parked, 2 redirected, 1 under construction, and 4 compromised websites, making a total of 17 websites ($34\%$) that are flagged out of the 50 randomly selected websites. 

\ignore{
\begin{figure*}[!t]
    \centering
\includegraphics[width=0.99\linewidth]
{figures/website_collage.png}
    \caption{Live screenshots of different websites from each cluster $C_1$ (green rectangle), $C_2$ (red rectangle), and $C_3$ (blue rectangle)}
    \label{fig:screen_shots}
\end{figure*}
}

\subsubsection{Website Contents \& Structure-based Analysis.}
\begin{figure}[!htbp]
    \centering
    \includegraphics[width=0.995\linewidth]{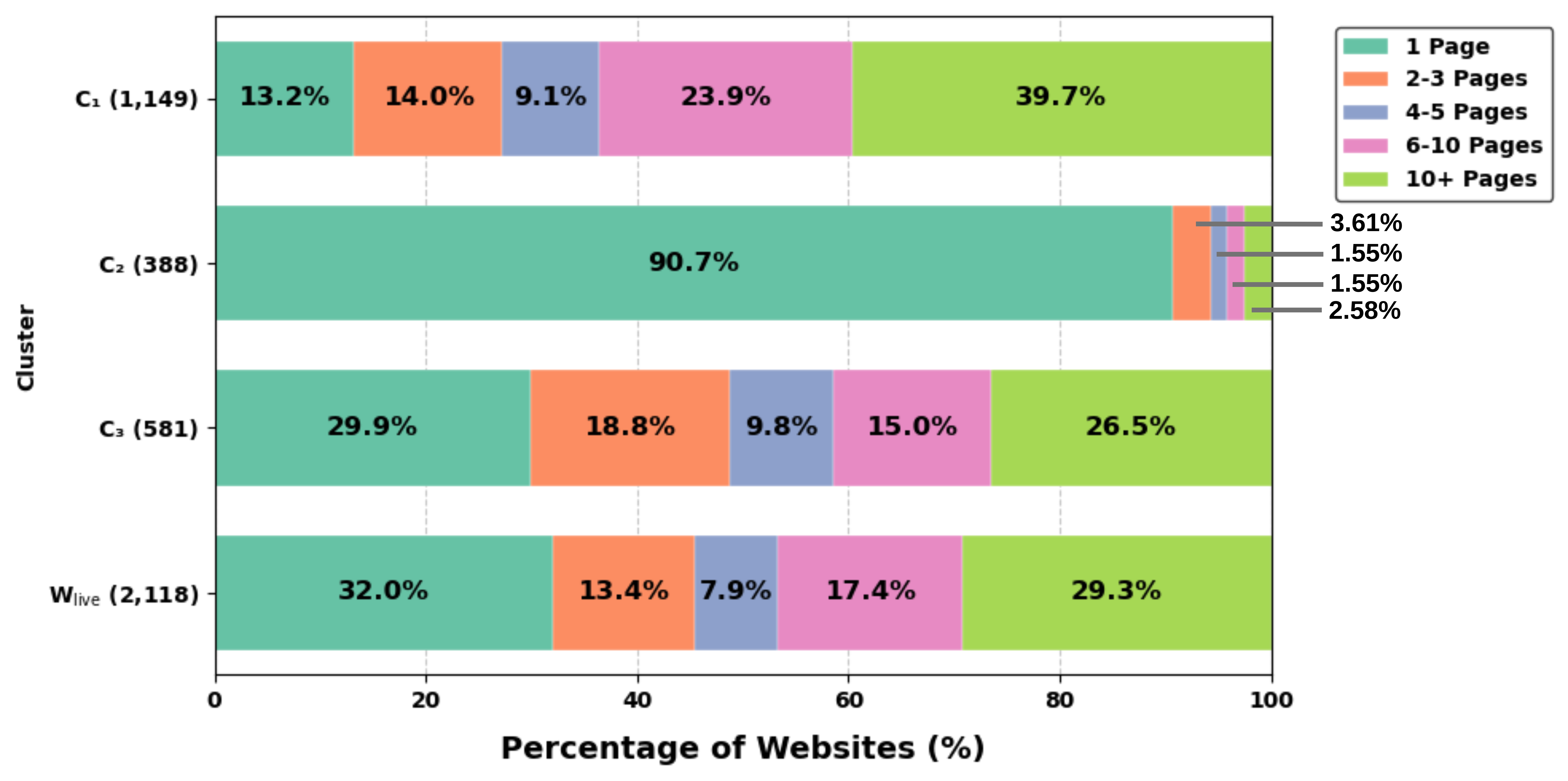}
    \caption{Websites' page-counts among different clusters}
    \label{fig:page_cnt}
\end{figure}

\begin{figure}[!h]
    \centering
    \includegraphics[width=0.75\columnwidth]{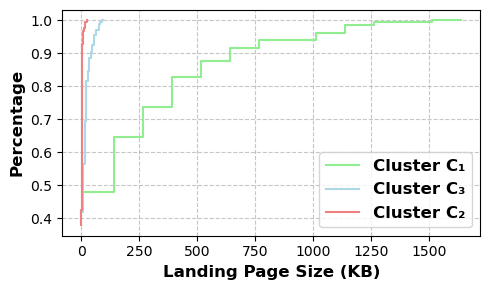}
    \caption{Cumulative Distribution Function (CDF) plot for the single-paged websites' landing page size}
    \label{fig:cdf_sing_page}
\end{figure}

First, we observe the page-count distribution of websites among the clusters as depicted in Figure \ref{fig:page_cnt}. 
This is also one of the most influential features for the malicious cluster $C_2$ as evident from the SHAP summary plot (Figure \ref{fig:exp_c2}). Additionally, we have encountered a notable number of parked websites in our manual analysis. Hence, we have investigated to confirm that the clustering is not biased only for this feature. We observe that cluster $C_2$ has on average the smallest {\em home page size} of $2.57$KB, where $90.7\%$ of the websites are single-paged. 
However, in cluster $C_1$ and $C_3$, we have only $13.23\%$ (152 out of 1,149 websites) and $29.95\%$ (174 out of 581 websites), single-paged websites, respectively. This indicates that even though $51.9\%$ of the total single-paged websites (352 out of 678 total single-paged websites) are mostly grouped in the malicious $C_2$ cluster, there are around $22.4\%$ and $25.7\%$ of total single-paged websites, respectively, in $C_1$ and $C_3$ clusters. 
The CDF plot in Figure \ref{fig:cdf_sing_page} also reveals that the size of the landing page follows a 
 very similar distribution in $C_2$ and $C_3$, with very minimal content on the landing page. On the other hand, almost $35\%$ of the single-paged websites in $C_1$ have a landing page size $\ge250KB$. 
Also, the websites residing in cluster $C_2$ have more frequent usage of `help', `war', `for\_to', `4\_2' words 
compared with the other 2 clusters. 


\begin{table*}[!htbp]
\caption{Distribution Of `Parked' Websites At Different Stages Among Clusters}
\centering
\resizebox{1.65\columnwidth}{!}{
\begin{tabular}{|p{2.10cm}|p{2.25cm}|p{2.35cm}|p{2.50cm}|p{2.50cm}|p{2.50cm}|}
\hline
 {\textbf{Cluster ID (Size)}} & {\textbf{\#parked sites in IQS}} & {\textbf{\#parked sites in FES}} & {\textbf{\#live sites in FES back from parked status}} & {\textbf{\# of live sites in IQS but gets parked in FES}} & {\textbf{\# of sites remained parked in both stages}}\\

 \hline
 
$C_1$(1,149) & 30(2.61\%) & 105(9.14\%) & 112(9.75\%) & 96(8.36\%) & 9(0.78\%)\\
\hline
$C_2$(388) & 129(33.25\%) & 127(32.73\%) & 28(7.22\%) & 117(30.15\%) & 10(2.58\%)\\
\hline
$C_3$(581) & 56(9.64\%) & 53(9.12\%) & 53(9.12\%) & 50(8.61\%) & 3(0.52\%)\\
\hline

$W_{live}$(2,118) & 215(10.15\%)  & 285(13.46\%) & 193(9.11\%) & 263(12.41\%) & 22(1.04\%)\\
\hline

\end{tabular}}

\label{table:parking_status}
\end{table*}

\begin{figure}[!htbp]
    \centering
    \includegraphics[width=0.85\columnwidth]{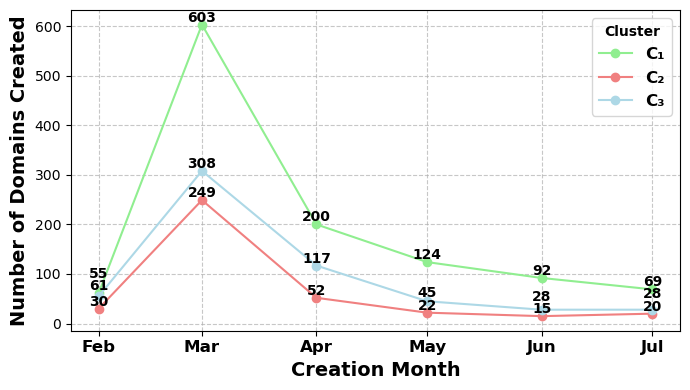}
    \caption{Domain creation trends by clusters (all $W_{live}$ Websites)}
    \label{fig:month_dom_creation_trend}
\end{figure}

\begin{figure}[!htbp]
    \centering
    \begin{subfigure}[b]{0.96\columnwidth}
        \centering
        \includegraphics[width=0.85\textwidth]{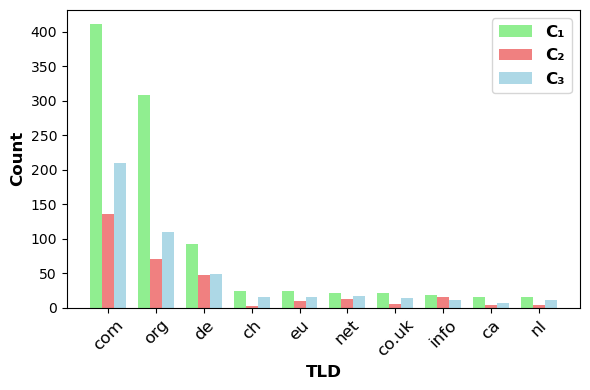}
        \vspace{-0.6em}
        \caption{Distribution of Top 10 TLDs across clusters}
        \label{fig:all_tld}
    \end{subfigure}
    \hfill 
    \begin{subfigure}[b]{0.96\columnwidth}
        \centering
        \includegraphics[width=0.85\textwidth]{figures/cc_TLD.png}   \caption{Distribution of ccTLDs across clusters with cost}
\label{fig:ccTLD}
    \end{subfigure}
    \caption{Analysis of various TLDs across clusters}
    \label{fig:tld_analysis_results}
\end{figure}

Additionally, we have provided the distribution of the parked websites across all 3 clusters for two stages of this study as highlighted in Table \ref{table:parking_status}. The two stages considered are the {\em IPQS query stage (IQS)} and the {\em feature extraction stage (FES)}. The table further illustrates that 28 websites ($7.2\%$) from cluster $C_2$ transitioned from a \textit{parking} state in the {\em IQS} stage to a \textit{live} state in the {\em FES} stage. Conversely, 117 websites ($30.2\%$) that were normal in the {\em IQS} stage got parked in the {\em FES} stage. 
This is an indication that the websites were either compromised or the owners of the websites were intentionally changing the structure of those websites periodically to evade detection. Moreover, Figure \ref{fig:month_dom_creation_trend} illustrates that the highest percentage of $C_2$ and $C_3$ websites are concentrated in March and April (2022), with a subsequent decline in May and June (2022). This pattern suggests a tendency for attackers to fish in troubled waters when very little information is out and more uncertainty is present on that particular event. Furthermore, 
Figure \ref{fig:tld_analysis_results}(a) shows that {\tt .com}, {\tt .org}, and {\tt .de} are the most used TLDs in all three clusters. Again, 
Figure \ref{fig:tld_analysis_results}(b) shows that
the {\tt .de} ccTLD is the most prevalent one (with a cheaper registration cost of USD 2.90 per year) along with other ccTLDs like {\tt .eu} (European Union),  {\tt .ch} (Switzerland), {\tt .nl} (Netherlands), {\tt .uk} (United Kingdom), {\tt .at} (Austria), {\tt .dk} (Denmark), and {\tt .fr} (France)
suggesting these websites are geo-located in European regions.

\noindent \textbf{Characterization Summary and Insights.} Our analysis on war-themed websites with both IPQS and manual investigation has led us to the following indicators of compromises: (i) presence of cheap TLDs like {\tt .de}, {\tt .us}, {\tt .pp.ua}, {\tt .live}; (ii) frequent use of hyphen character; (iii) longer domain length; (iv) higher Spamhaus badness score for the associated TLD; (v) presence of personal bank account in donation; (vi) smaller home page size (KBs); (vii) lower unique page counts. (viii) frequent use of event keywords in both the domain and website's contents, which answers \textbf{RQ3}. Furthermore, we provide a summary of characterization for 3 clusters, which can be used by analysts for decision-making on {\em course-of-actions} by analysts.  

 \noindent \textbf{Higher percentage of benign websites in cluster $C_1$.} We have found normal behavior in most website contents, and many are not even related to the war scenarios (i.e., collected due to their domain name). For the donation websites in $C_1$, the recipient accounts are not linked with personal bank accounts. There are variations in website structures, and almost all the websites are crafted carefully for legitimate business purposes with well-structured pages. 
    
  \noindent \textbf{Higher percentages of malicious websites in cluster $C_2$.} One of the most interesting finds for this group of websites is that $90.7\%$ of these are single-paged websites. This behavior is an indication that the attacker's strategy is not to invest much time in creating more complete and structured websites but to propagate the malicious websites to a large number of recipients in the early phase of the event. 
  The five discovered scam campaigns are namely-- donation, charity, crypto, publicity, and service scams that employ distinct strategic approaches to enhance their effectiveness in achieving malicious objectives. 
  Also, the usage of cheap TLDs, displaying of so far collected money to gain trust, and oscillating between parked and normal state -- these characteristics make this cluster a potential malicious group for fraud and scam campaign sites (e.g., fake donations), which answers \textbf{RQ4}. 
  
  \noindent \textbf{Mostly benign websites in cluster $C_3$.} Websites from this cluster also express normal behavior, except we have found some websites that have too many outgoing links, which are mostly social media links. This indicates that websites from this cluster were probably used to gain attention from the outside world about the war situation with social media campaigns. We have also encountered some of the donation campaigns among this cluster with fewer cases of compromised, parked, and redirected websites.

%% file: discussion.tex
Although we have provided only one case study based on war-themed websites, our methodology can be applied to characterize other event-themed malicious website campaigns. 
Moreover, our methodology can provide valuable insights to defenders, as they can group websites and only need to investigate a subset of the websites, effectively reducing the analysis time. In addition, the characterization process can reveal deeper insights about each cluster, potentially accelerating the mitigation of these attacks (e.g., taking down or blocking websites). The explanation scheme 
reveals why a particular instance belongs to a certain cluster (i.e., local explanation), enhancing the confidence of the defenders to take action. 

Still, the study has some limitations.
First, we focus on studying event-themed campaigns at their early stage, without quantifying the notion of {\em early stage} (e.g., should the first week of a campaign be considered early stage?). This issue is important because datasets of an event-themed malicious web campaign would evolve with time, and this early stage may also vary based on the gravity and scale of the event. 
Second, the website domain names in our dataset only contain the words `Ukraine' or `Ukrainian', but not necessarily `Russia' or `Russian'. This means that the dataset may have missed other relevant domains. 
Third, our manual analysis to understand the performance of the methodology is conducted on a small number of websites (i.e., 50 websites from each cluster) because there is no reliable ground-truth data available. 
Fourth, we do not investigate whether any of the themed websites used cloaking techniques \cite{CrawlPhish_Oest_sp22_cloaking}, which means that we may have missed some malicious websites.
Fifth, in the feature extraction process, we only consider English-based keywords. This means that we have overlooked relevant keywords in other languages (e.g., Russian, Ukrainian, or other languages in the EU region). Sixth, our study shows that only 11.56\% of the websites (i.e., 2,118 out of 18,321) were live when we initiated the investigation (2024), meaning a large portion of malicious websites have already been taken down.
Conducting the investigation at an earlier time (i.e., closer to the event start date) would have revealed more information about the themed websites.

%% file: conclusion.tex
Event-themed malicious web campaigns are an emergent threat. To address the problem, we propose a methodology to characterize them via explainable unsupervised clustering because these attacks are typically new, and labeled data is often not available for training supervised models. 
Our case study with the Russia-Ukraine war-themed websites shows indicators of these attacks, such as the presence of cheap TLDs, lower page counts, and the presence of personal bank accounts. 
Our study also reveals that a particular cluster has a higher rate of hosting malicious websites than others.

In addition to addressing the aforementioned limitations, there are interesting open problems. One is to investigate 
coordinated event-themed malicious campaigns. 
Another is to investigate the payloads (e.g., attachments) of event-themed web campaigns in waging malware or ransomware attacks. 
It is also interesting to investigate deep unsupervised learning (e.g., auto-encoders) and semi-supervised learning, rather than unsupervised clustering, in the context of the methodology.
Moreover, it is interesting to conduct case studies on other kinds of even-themed attacks, such as disaster-themed, humanitarian-cause themed, political-conflict-themed, and severe-climate-change-themed.
Furthermore, it is interesting to investigate the evasion techniques (if any) that have been used by event-themed attacks.
Last but not least, the community should investigate the {\em psychological} sophistication of these attacks \cite{montanez2023Quantifying,XuPIEEE2024} and develop cyber social engineering kill chains 
\cite{RosaSocialEngineeringKillChainSciSec2022} for different event-themed cyber threats. 

%% file: conference_101719.bbl
\begin{thebibliography}{10}
\providecommand{\url}[1]{#1}
\csname url@samestyle\endcsname
\providecommand{\newblock}{\relax}
\providecommand{\bibinfo}[2]{#2}
\providecommand{\BIBentrySTDinterwordspacing}{\spaceskip=0pt\relax}
\providecommand{\BIBentryALTinterwordstretchfactor}{4}
\providecommand{\BIBentryALTinterwordspacing}{\spaceskip=\fontdimen2\font plus
\BIBentryALTinterwordstretchfactor\fontdimen3\font minus \fontdimen4\font\relax}
\providecommand{\BIBforeignlanguage}[2]{{%
\expandafter\ifx\csname l@#1\endcsname\relax
\typeout{** WARNING: IEEEtran.bst: No hyphenation pattern has been}%
\typeout{** loaded for the language `#1'. Using the pattern for}%
\typeout{** the default language instead.}%
\else
\language=\csname l@#1\endcsname
\fi
#2}}
\providecommand{\BIBdecl}{\relax}
\BIBdecl

\bibitem{al2022covid}
A.~F. Al-Qahtani and S.~Cresci, ``The covid-19 scamdemic: A survey of phishing attacks and their countermeasures during covid-19,'' \emph{IET Information Security}, vol.~16, no.~5, pp. 324--345, 2022.

\bibitem{mir2020characterizing}
M.~M.~A. Pritom, K.~M. Schweitzer, R.~M. Bateman, M.~Xu, and S.~Xu, ``Characterizing the landscape of covid-19 themed cyberattacks and defenses,'' in \emph{2020 IEEE International Conference on Intelligence and Security Informatics (ISI)}, 2020, pp. 1--6.

\bibitem{khatri2023global}
S.~Khatri, A.~K. Cherukuri, and F.~Kamalov, ``Global pandemics influence on cyber security and cyber crimes,'' \emph{arXiv preprint arXiv:2302.12462}, 2023.

\bibitem{rakesh_verma_phishing_with_disaster}
R.~Verma, D.~Crane, and O.~Gnawali, ``Phishing during and after disaster: Hurricane harvey,'' in \emph{2018 Resilience Week (RWS)}, 2018, pp. 88--94.

\bibitem{ukraine_cyber_scams}
MarCom, ``War in ukraine sparks new cyber scams,'' https://ndbt.com/insights/war-in-ukraine-sparks-new-cyber-scams/, 2022, (Accessed \ on \ 5 \ April, \ 2024).

\bibitem{ukraine_charity_scams}
DISB, ``Beware of ukrainian crisis charity scams,'' https://disb.dc.gov/page/beware-ukrainian-crisis-charity-scams, 2022, (Accessed \ on \ 5 \ April, \ 2024).

\bibitem{XuPIEEE2024}
T.~T. Longtchi, R.~M. Rodriguez, L.~Al{-}Shawaf, A.~Atyabi, and S.~Xu, ``Internet-based social engineering psychology, attacks, and defenses: {A} survey,'' \emph{Proc. {IEEE}}, vol. 112, no.~3, pp. 210--246, 2024.

\bibitem{Oest_20_apwgSymp_scamPandemic}
M.~Bitaab, H.~Cho, A.~Oest, P.~Zhang, Z.~Sun, R.~Pourmohamad, D.~Kim, T.~Bao, R.~Wang, Y.~Shoshitaishvili, A.~Doupé, and G.-J. Ahn, ``Scam pandemic: How attackers exploit public fear through phishing,'' in \emph{2020 APWG Symposium on Electronic Crime Research (eCrime)}, 2020, pp. 1--10.

\bibitem{mir_covid_website20}
M.~M.~A. Pritom, K.~M. Schweitzer, R.~M. Bateman, M.~Xu, and S.~Xu, ``Data-driven characterization and detection of covid-19 themed malicious websites,'' in \emph{2020 IEEE International Conference on Intelligence and Security Informatics (ISI)}, 2020, pp. 1--6.

\bibitem{Behzad_2023CovidScams_AsiaCCS}
\BIBentryALTinterwordspacing
B.~Ousat, M.~A. Tofighi, and A.~Kharraz, ``An end-to-end analysis of covid-themed scams in the wild,'' in \emph{Proceedings of the 2023 ACM Asia Conference on Computer and Communications Security}, ser. ASIA CCS '23.\hskip 1em plus 0.5em minus 0.4em\relax New York, NY, USA: Association for Computing Machinery, 2023, p. 509–523. [Online]. Available: \url{https://doi.org/10.1145/3579856.3582831}
\BIBentrySTDinterwordspacing

\bibitem{krichen2023managing}
M.~Krichen, M.~S. Abdalzaher, M.~Elwekeil, and M.~M. Fouda, ``Managing natural disasters: An analysis of technological advancements, opportunities, and challenges,'' \emph{Internet of Things and Cyber-Physical Systems}, 2023.

\bibitem{magafasrussia}
L.~Magafas and K.~Demertzis, ``Russia vs ukraine cyberwarfare: Lessons learned.''

\bibitem{mohee2022cyber}
A.~Mohee, ``Cyber war: The hidden side of the russian-ukrainian crisis,'' 2022.

\bibitem{vu2022getting}
A.~V. Vu, D.~R. Thomas, B.~Collier, A.~Hutchings, R.~Clayton, and R.~Anderson, ``Getting bored of cyberwar: Exploring the role of civilian participation in the russia-ukraine cyber conflict,'' 2022.

\bibitem{gabrian2022russia}
C.-A. GABRIAN, ``How the russia-ukraine war may change the cybercrime ecosystem,'' \emph{BULLETIN OF" CAROL I" NATIONAL DEFENCE UNIVERSITY}, vol.~11, no.~4, pp. 43--49, 2022.

\bibitem{PhishingURL_data_diversity_Nsys2024_mir}
\BIBentryALTinterwordspacing
M.~Mia, D.~Derakhshan, and M.~M.~A. Pritom, ``Can features for phishing url detection be trusted across diverse datasets? a case study with explainable ai,'' in \emph{Proceedings of the 11th International Conference on Networking, Systems, and Security}, ser. NSysS '24.\hskip 1em plus 0.5em minus 0.4em\relax New York, NY, USA: Association for Computing Machinery, 2025, p. 137–145. [Online]. Available: \url{https://doi.org/10.1145/3704522.3704532}
\BIBentrySTDinterwordspacing

\bibitem{sensors23_malURLs_ml}
\BIBentryALTinterwordspacing
S.~Abad, H.~Gholamy, and M.~Aslani, ``Classification of malicious urls using machine learning,'' \emph{Sensors}, vol.~23, no.~18, 2023. [Online]. Available: \url{https://www.mdpi.com/1424-8220/23/18/7760}
\BIBentrySTDinterwordspacing

\bibitem{phish_urls_verma_codaspy15}
\BIBentryALTinterwordspacing
R.~Verma and K.~Dyer, ``On the character of phishing urls: Accurate and robust statistical learning classifiers,'' in \emph{Proceedings of the 5th ACM Conference on Data and Application Security and Privacy}, ser. CODASPY '15.\hskip 1em plus 0.5em minus 0.4em\relax New York, NY, USA: Association for Computing Machinery, 2015, p. 111–122. [Online]. Available: \url{https://doi.org/10.1145/2699026.2699115}
\BIBentrySTDinterwordspacing

\bibitem{xu13_crosslayer_malwebsite}
\BIBentryALTinterwordspacing
L.~Xu, Z.~Zhan, S.~Xu, and K.~Ye, ``Cross-layer detection of malicious websites,'' in \emph{Proceedings of the Third ACM Conference on Data and Application Security and Privacy}, ser. CODASPY '13.\hskip 1em plus 0.5em minus 0.4em\relax New York, NY, USA: Association for Computing Machinery, 2013, p. 141–152. [Online]. Available: \url{https://doi.org/10.1145/2435349.2435366}
\BIBentrySTDinterwordspacing

\bibitem{ccs21_phishing_https_certificate}
\BIBentryALTinterwordspacing
D.~Kim, H.~Cho, Y.~Kwon, A.~Doup\'{e}, S.~Son, G.-J. Ahn, and T.~Dumitras, ``Security analysis on practices of certificate authorities in the https phishing ecosystem,'' in \emph{Proceedings of the 2021 ACM Asia Conference on Computer and Communications Security}, ser. ASIA CCS '21.\hskip 1em plus 0.5em minus 0.4em\relax New York, NY, USA: Association for Computing Machinery, 2021, p. 407–420. [Online]. Available: \url{https://doi.org/10.1145/3433210.3453100}
\BIBentrySTDinterwordspacing

\bibitem{phishpatterns_IMC22_Perdisci}
\BIBentryALTinterwordspacing
K.~Subramani, W.~Melicher, O.~Starov, P.~Vadrevu, and R.~Perdisci, ``Phishinpatterns: measuring elicited user interactions at scale on phishing websites,'' in \emph{Proceedings of the 22nd ACM Internet Measurement Conference}, ser. IMC '22.\hskip 1em plus 0.5em minus 0.4em\relax New York, NY, USA: Association for Computing Machinery, 2022, p. 589–604. [Online]. Available: \url{https://doi.org/10.1145/3517745.3561467}
\BIBentrySTDinterwordspacing

\bibitem{xu14_cns_evasion_malwebsite}
L.~Xu, Z.~Zhan, S.~Xu, and K.~Ye, ``An evasion and counter-evasion study in malicious websites detection,'' in \emph{2014 IEEE Conference on Communications and Network Security}, 2014, pp. 265--273.

\bibitem{goodfellow2015explainingICLR2015}
I.~J. Goodfellow, J.~Shlens, and C.~Szegedy, ``Explaining and harnessing adversarial examples,'' in \emph{3rd International Conference on Learning Representations (ICLR'2015)}, Y.~Bengio and Y.~LeCun, Eds., 2015.

\bibitem{das2019sok_rakesh}
A.~Das, S.~Baki, A.~El~Aassal, R.~Verma, and A.~Dunbar, ``Sok: a comprehensive reexamination of phishing research from the security perspective,'' \emph{IEEE Communications Surveys \& Tutorials}, vol.~22, no.~1, pp. 671--708, 2019.

\bibitem{song2021advanced}
F.~Song, Y.~Lei, S.~Chen, L.~Fan, and Y.~Liu, ``Advanced evasion attacks and mitigations on practical ml-based phishing website classifiers,'' \emph{International Journal of Intelligent Systems}, vol.~36, no.~9, pp. 5210--5240, 2021.

\bibitem{Oest20_Phishtime}
\BIBentryALTinterwordspacing
A.~Oest, Y.~Safaei, P.~Zhang, B.~Wardman, K.~Tyers, Y.~Shoshitaishvili, and A.~Doup{\'e}, ``{PhishTime}: Continuous longitudinal measurement of the effectiveness of anti-phishing blacklists,'' in \emph{29th USENIX Security Symposium (USENIX Security 20)}.\hskip 1em plus 0.5em minus 0.4em\relax USENIX Association, Aug. 2020, pp. 379--396. [Online]. Available: \url{https://www.usenix.org/conference/usenixsecurity20/presentation/oest-phishtime}
\BIBentrySTDinterwordspacing

\bibitem{CrawlPhish_Oest_sp22_cloaking}
P.~Zhang, A.~Oest, H.~Cho, Z.~Sun, R.~Johnson, B.~Wardman, S.~Sarker, A.~Kapravelos, T.~Bao, R.~Wang, Y.~Shoshitaishvili, A.~Doupe, and G.~Ahn, ``Crawlphish: Large-scale analysis of client-side cloaking techniques in phishing,'' \emph{IEEE Security \& Privacy}, vol.~20, no.~02, pp. 10--21, mar 2022.

\bibitem{Oest_Phishfarm_sp19_blacklist}
A.~Oest, Y.~Safaei, A.~Doupé, G.-J. Ahn, B.~Wardman, and K.~Tyers, ``Phishfarm: A scalable framework for measuring the effectiveness of evasion techniques against browser phishing blacklists,'' in \emph{2019 IEEE Symposium on Security and Privacy (SP)}, 2019, pp. 1344--1361.

\bibitem{Oest_sp23_BeyondPhish}
M.~Bitaab, H.~Cho, A.~Oest, Z.~Lyu, W.~Wang, J.~Abraham, R.~Wang, T.~Bao, Y.~Shoshitaishvili, and A.~Doupé, ``Beyond phish: Toward detecting fraudulent e-commerce websites at scale,'' in \emph{2023 IEEE Symposium on Security and Privacy (SP)}, 2023, pp. 2566--2583.

\bibitem{oest2022neutralizing}
A.~Oest, P.~Zhang, and R.~Johnson, ``Neutralizing evasion techniques of malicious websites,'' Apr.~28 2022, uS Patent App. 17/079,190.

\bibitem{Spamhaus}
T.~S.~P. SLU, ``The world's most abused domain registrars,'' https://www.spamhaus.org/statistics/registrars/, accessed on 13 March, 2024.

\bibitem{benesty2009pearson}
J.~Benesty, J.~Chen, Y.~Huang, and I.~Cohen, ``Pearson correlation coefficient,'' in \emph{Noise reduction in speech processing}.\hskip 1em plus 0.5em minus 0.4em\relax Springer, 2009, pp. 37--40.

\bibitem{IKOTUN2023178}
\BIBentryALTinterwordspacing
A.~M. Ikotun, A.~E. Ezugwu, L.~Abualigah, B.~Abuhaija, and J.~Heming, ``K-means clustering algorithms: A comprehensive review, variants analysis, and advances in the era of big data,'' \emph{Information Sciences}, vol. 622, pp. 178--210, 2023. [Online]. Available: \url{https://www.sciencedirect.com/science/article/pii/S0020025522014633}
\BIBentrySTDinterwordspacing

\bibitem{kmedoid_park2009simple}
H.-S. Park and C.-H. Jun, ``A simple and fast algorithm for k-medoids clustering,'' \emph{Expert systems with applications}, vol.~36, no.~2, pp. 3336--3341, 2009.

\bibitem{nielsen2016hierarchical}
F.~Nielsen and F.~Nielsen, ``Hierarchical clustering,'' \emph{Introduction to HPC with MPI for Data Science}, pp. 195--211, 2016.

\bibitem{weber2022gaussian}
C.~M. Weber, D.~Ray, A.~A. Valverde, J.~A. Clark, and K.~S. Sharma, ``Gaussian mixture model clustering algorithms for the analysis of high-precision mass measurements,'' \emph{Nuclear Instruments and Methods in Physics Research Section A: Accelerators, Spectrometers, Detectors and Associated Equipment}, vol. 1027, p. 166299, 2022.

\bibitem{Shillouette}
K.~R. Shahapure and C.~Nicholas, ``Cluster quality analysis using silhouette score,'' in \emph{2020 IEEE 7th International Conference on Data Science and Advanced Analytics (DSAA)}, 2020, pp. 747--748.

\bibitem{silhouttee_just_1_januzaj2023determining}
Y.~Januzaj, E.~Beqiri, and A.~Luma, ``Determining the optimal number of clusters using silhouette score as a data mining technique.'' \emph{International Journal of Online \& Biomedical Engineering}, vol.~19, no.~4, 2023.

\bibitem{lundberg2017unified}
S.~M. Lundberg and S.-I. Lee, ``A unified approach to interpreting model predictions,'' \emph{Advances in neural information processing systems}, vol.~30, 2017.

\bibitem{unservised_explanation}
\BIBentryALTinterwordspacing
A.~Morichetta, P.~Casas, and M.~Mellia, ``Explain-it: Towards explainable ai for unsupervised network traffic analysis,'' in \emph{Proceedings of the 3rd ACM CoNEXT Workshop on Big DAta, Machine Learning and Artificial Intelligence for Data Communication Networks}, ser. Big-DAMA '19.\hskip 1em plus 0.5em minus 0.4em\relax New York, NY, USA: Association for Computing Machinery, 2019, p. 22–28. [Online]. Available: \url{https://doi.org/10.1145/3359992.3366639}
\BIBentrySTDinterwordspacing

\bibitem{pritom_law_support_cns22}
M.~M.~A. Pritom and S.~Xu, ``Supporting law-enforcement to cope with blacklisted websites: Framework and case study,'' in \emph{2022 IEEE Conference on Communications and Network Security (CNS)}, 2022, pp. 181--189.

\bibitem{IPQualityScore}
\BIBentryALTinterwordspacing
IPQS. (2024) Ip quality score. [Online]. Available: \url{https://www.ipqualityscore.com}
\BIBentrySTDinterwordspacing

\bibitem{lallie2020cyber_COVID19Paper}
H.~S. Lallie, L.~A. Shepherd, J.~R.~C. Nurse, A.~Erola, G.~Epiphaniou, C.~Maple, and X.~Bellekens, ``Cyber security in the age of covid-19: A timeline and analysis of cyber-crime and cyber-attacks during the pandemic,'' 2020.

\bibitem{COVID19_data_checkphish}
CheckPhish, ``Covid-19 (coronavirus) phishing \& scam tracker,'' https://checkphish.ai/coronavirus-scams-tracker, 2020, accessed on 15 May, 2020.

\bibitem{calinski1974dendrite}
T.~Cali{\'n}ski and J.~Harabasz, ``A dendrite method for cluster analysis,'' \emph{Communications in Statistics-theory and Methods}, vol.~3, no.~1, pp. 1--27, 1974.

\bibitem{akaike2011akaike}
H.~Akaike, ``Akaike’s information criterion,'' \emph{International encyclopedia of statistical science}, pp. 25--25, 2011.

\bibitem{montanez2023Quantifying}
R.~Monta{\~n}ez, T.~Longtchi, K.~Gwartney, E.~Ear, D.~Azari, C.~Kelley, and S.~Xu, ``Quantifying psychological sophistication of malicious emails,'' in \emph{Proceedings of International Conference on Science of Cyber Security (SciSec'2023)}, 2023, pp. 319--331.

\bibitem{RosaSocialEngineeringKillChainSciSec2022}
R.~M. Rodriguez and S.~Xu, ``Cyber social engineering kill chain,'' in \emph{Proceedings of International Conference on Science of Cyber Security (SciSec'2022)}, 2022, pp. 487--504.

\end{thebibliography}
